\documentclass[acmsmall, nonacm]{acmart}

\usepackage{subcaption}
\usepackage[table]{xcolor}
\usepackage{pifont}
\usepackage{wasysym}
\usepackage{threeparttable}
\usepackage{algorithm}
\usepackage{algpseudocode}

\renewcommand\footnotetextcopyrightpermission[1]{}

\AtBeginDocument{%
  }

\begin{document}

\title{zQR: A Verifiable QR-Driven zkSNARK Proof Verification Framework for Mobile Platforms}

\author{Goshgar C. Ismayilov}
\email{goshgar.ismayilov@boun.edu.tr}

\renewcommand{\shortauthors}{Ismayilov et al.}

\begin{abstract}
Privacy is one of the fundamental rights of individuals in modern societies. Yet, the practical adoption of privacy-preserving technologies in daily interactions remains limited. Zero-knowledge proofs offer strong privacy guarantees but are often hindered by their technical complexity. In this paper, we advance the idea of \textit{verifiable QR codes} that enable off-line verifiers to verify proofs encoded in QR codes. Based on this core idea, we build a novel QR-driven zkSNARK proof verification framework (i.e., \textit{zQR}) for mobile platforms. The framework integrates blockchain for auditability, non-repudiation and logging; and large-language models for automatic circuit generation. We perform a security discussion of the framework by considering multiple attack surfaces. Furthermore, we present an experimental evaluation measuring temporal costs (proof generation and verification latency, QR code encoding and decoding latency) and financial costs (blockchain gas consumption). Our results demonstrate the feasibility of \textit{zQR} as a proof-of-concept framework for privacy-preserving verification on mobile platform where proofs are compactly represented with QR code symbol version of 19 with low error correction level. Finally, we discuss potential applications, current limitations and future directions for the broader adoption of privacy-preserving technologies in daily interactions.
\end{abstract}

\maketitle

\section{Introduction}

Privacy is a fundamental right of individuals to control how their sensitive data are collected, used, shared and disclosed. Violation of privacy has the potential to have catastrophic consequences, especially in modern societies. In the past, many global data breach and leakage incidents occurred, affecting millions of individuals (e.g., the Yahoo case in 2014 \cite{yahoo_breach} that affected over 500 million accounts). The adoption of privacy-preserving techniques in daily life may be among the best precautions against such malicious threats. The literature addresses privacy in numerous ways including trusted execution environments \cite{guo2025}, zero-knowledge proofs \cite{sun2021} and secure multi-party computation \cite{zhou2021}. Specifically, advancements in zero-knowledge proof have been recently accelerated in terms of novel protocols (e.g., zkSNARKs \cite{zksnarks}, zkSTARKs \cite{zkstarks}, Bulletproofs \cite{bulletproofs}, Plonk \cite{plonk}) as well as toolchains (e.g., Circom \cite{belles2022}, ZoKrates \cite{eberhardt2018}). However, the merits of zero-knowledge proof have been accompanied by their own challenges, most notably the technical complexity of its adoption. On the opposite side, QR codes have experienced widespread use since they require no specialized hardware, operate offline, are widely supported by smartphones and provide a familiar interaction model. In this work, we address the adoption issue of zero-knowledge proof for daily applications with the help of verifiable QR codes.

\textit{Privacy-driven motivations.} Modern data protection regulations, most notably the General Data Protection Regulation (GDPR), stress the following three core principles. \textit{Data minimization} refers to the collection of only the personal data that is strictly necessary for a specific purpose (e.g., verifying that an individual is over 18). \textit{Selective disclosure} enables individuals to reveal only specific attributes or statements about their data (e.g., proving possession of a valid travel ticket without revealing transaction details). \textit{Privacy by design} suggests directly embedding privacy protections into architectural designs (e.g., designing a zero-knowledge proof-based mobile authentication application). Despite all these principles in theory, there still remains a lack of usable and efficient privacy-preserving systems in practice.

\textit{Accessibility-driven motivations.} In the literature, there are works that propose privacy-preserving and secure protocols for a wide range of critical problems including financial payments \cite{ismayilov20252}, healthcare \cite{cao2022}, insurance \cite{itanyi2023}, voting \cite{wu2023} and supply chain \cite{heiss2023}. However, most of these works primarily focus on their own theoretical and empirical contributions and discuss their real-world adoption via strongly-restricted assumptions. This naturally widens the gap between the theoretical improvements of digital zero-knowledge proofs and their practical reflections over time. In this work, we attempt to narrow this gap by leveraging the popularity of QR codes in daily life. We anticipate that such attempts reveal more potential use cases for adoption and contribute toward a more privacy-preserving but publicly verifiable and auditable world.

\textit{Literature-driven motivations.} The integration of QR codes into zero-knowledge proof verification has not been well-studied in the literature. Although there exist a limited number of prior works addressing this issue, no systematic study currently explores verifiable QR-driven zkSNARK verification on mobile platforms. The absence of any standardization on this issue also hinders reproducibility, comparative analysis and performance optimization across different implementations. This paper aims to fill this gap by proposing the \textit{zQR} framework.

\textit{Challenges.} To build a verifiable QR code framework through zero-knowledge proof on mobile platforms, several critical challenges must be properly addressed. \textit{Lightweightness}: mobile devices often have 
limited capacity in terms of computation, memory and battery to perform complex and energy-intensive operations. \textit{Latency}: a real-time system must have instant or near-instant responses for a better user experience. \textit{Scalability}: the increasing number of users and blockchain transactions must be effectively supported. \textit{Privacy-preservation}: minimizing data disclosure and protecting the privacy of sensitive data must be among the main objectives. \textit{Low-effort}: end-users must not require a complex technical background in zero-knowledge proof. \textit{Security}: the framework must be resistant to threats coming from various attack surfaces.

\textit{Contributions.} The primary contributions of this work are listed as follows:
\begin{itemize}
\itemsep0em
\item[$\checkmark$] \textit{Novelty and Core Contribution.} This work advances the idea of verifiable QR codes for a more privacy-preserving but publicly verifiable and auditable world with the adoption of zero-knowledge proofs on blockchain. 

\item[$\checkmark$] \textit{Novel Framework.} This work proposes a novel QR-driven zkSNARK proof verification framework (\textit{zQR}) for mobile platforms. It explores QR-based proof verification within a unified architectural framework. The framework integrates large-language models for low-code proof circuit generation and blockchain for on-chain proof verification.

\item[$\checkmark$] \textit{Experimental Analysis.} This work performs an experimental study measuring performance in terms of temporal cost (for proof generation and verification and QR code encoding and decoding) and financial cost (for smart contract deployment and execution on a public blockchain testnet). The work also discusses the \textit{zQR} security from the perspective of zero-knowledge proof, QR code and blockchain threat surfaces.

\item[$\checkmark$] \textit{Applications, Limitations and Future Directions.} This work discusses the potential use cases of the \textit{zQR} framework across several real-world application domains. Moreover, the work clearly identifies the current limitations of the framework and provides a collection of future directions to promote potential advancements in the field.
\end{itemize}

\section{Related Work}

In the literature, very few scientific works address verifiable QR codes through zero-knowledge proof. The work \cite{gokulakrishnan2025} recently proposes a zero-knowledge proof-based product source verification system in supply chain where third-party verifiers rely on QR code-based data authentication. This system uses zkSNARKs for off-chain proof generation and blockchain for on-chain proof verification. However, their system differs from ours in several key aspects. First, QR codes in their design are not inherently verifiable. Rather, they redirect users to a server where verification occurs, which prevents the system from operating off-line. Second, the proof construction is application-specific and non-customizable. In contrast, our approach enables users to dynamically and automatically generate arbitrary proof circuits, making the framework adaptable to a broader range of applications. The work \cite{gantait2022} discusses zkSNARK proof verification in the format of verifiable QR codes for voting systems where a polling officer scans and verifies the proof to confirm the eligibility of voters. Similar to the previous approach, their scheme is also application-specific and does not provide a systematic architectural framework or a comprehensive security discussion.

Among the most closely related works, the work \cite{zhou2024} recently proposes a more generic verifiable QR code scheme for mobile platforms based on the Plonk protocol \cite{plonk} and Verkle trees \cite{kuszmaul2019} 
where a verifier scans QR codes that users generate for their attributes. The scheme also integrates blockchain for contract-based on-chain proof verification. However, it remains tightly coupled to a verifiable anonymous credential construction based on Verkle trees and lacks a comprehensive experimental evaluation across different circuit configurations. The work \cite{de2022} aims to 
transfer verifiable credentials via QR codes for privacy-preserving vaccination passes. Hence, it is application-specific and provides no generalizable framework. Overall, the literature currently lacks a verifiable QR code framework that simultaneously integrates blockchain for on-chain verification and large-language models for automatic circuit generation while remaining application-agnostic. Table~\ref{table:related} summarizes the comparison among existing works and the \textit{zQR} framework.

\begin{table*}[htbp]
\caption{Comparison of Previous Studies on QR Codes for ZKP}
\footnotesize
\begin{center}
\begin{threeparttable}
\begin{tabular}{lcllcccc}
\hline
\cellcolor{gray!50} \textbf{Study} & 
\cellcolor{gray!50} \textbf{Year} & 
\cellcolor{gray!50} \textbf{Protocol} & 
\cellcolor{gray!50} \textbf{Application} & 
\cellcolor{gray!50} \textbf{Dynamic} & 
\cellcolor{gray!50} \textbf{LLM} & 
\cellcolor{gray!50} \textbf{Blockchain} & 
\cellcolor{gray!50} \textbf{Analysis} \\   
\hline
\cite{zhou2024} & 2024 & Plonk & Generic & \checkmark & 
$\times$ & \checkmark & $\LEFTcircle$ \\
\cite{gokulakrishnan2025} & 2025 & zkSNARKs & Supply Chain &  $\times$ & $\times$ & \checkmark & $\LEFTcircle$ \\
\cite{de2022} & 2022 & Hyperledger AnonCreds & Healthcare & $\times$ & $\times$ & 
\checkmark & $\Circle$ \\
\cite{gantait2022} & 2022 & zkSNARKs & Voting & $\times$ & 
$\times$ & \checkmark & $\Circle$ \\
\rowcolor{blue!20}
\textit{zQR} & 2026 & zkSNARKs & Generic & \checkmark & 
\checkmark & \checkmark & $\LEFTcircle$ \\
\hline
\end{tabular}
\begin{tablenotes}
\footnotesize
\item $\Circle$: No security threat discussion
\item $\LEFTcircle$: Informal security threat discussion
\item $\CIRCLE$: Formal security analysis
\end{tablenotes}
\end{threeparttable}
\label{table:related}
\end{center}
\end{table*}

\section{\textit{zQR} Framework}

The proposed \textit{zQR} framework involves three active roles as \textit{issuer}, \textit{prover} and \textit{verifier}; and two passive roles as \textit{smart contract} and \textit{large-language model}. Active 
roles are the entities that are able to initiate actions while passive roles are the entities that respond to requests upon invocation. Their responsibilities are defined as follows:

\begin{itemize}
\itemsep0em
\item \textit{Issuer} $(\mathcal{S})$. constructs proof circuits either manually or automatically with the assistance of large-language models; performs trusted setup to generate proving and verification keys for proof generation and verification respectively; and deploys proof-verifying 
smart contracts to the blockchain.

\item \textit{Prover} $(\mathcal{P})$. owns the proving key and certain sensitive information (e.g., age or health data); generates zero-knowledge proofs without disclosing private information; encodes proofs into verifiable QR codes for off-chain verification; and submits proofs to smart contracts for on-chain verification.

\item \textit{Verifier} $(\mathcal{V})$. owns the verification key and only public information (e.g., public commitments); decodes verifiable QR codes into corresponding proofs; and verifies proofs off-chain.

\item \textit{Smart Contract} $(\mathsf{SC})$. is automatically generated by embedding the verification key into the ZoKrates contract template; deployed to the blockchain by the issuer; and used to verify proofs on-chain upon request.

\item \textit{Large-Language Model} $(\mathsf{LLM})$. is prompted with a combination of a user prompt containing proof specifications and example inputs; and a system prompt containing instructions, boundaries and constraints; and returns a proof circuit to the issuer.
\end{itemize}

\begin{figure*}[htbp]
\centerline{\includegraphics[width=0.99\textwidth]{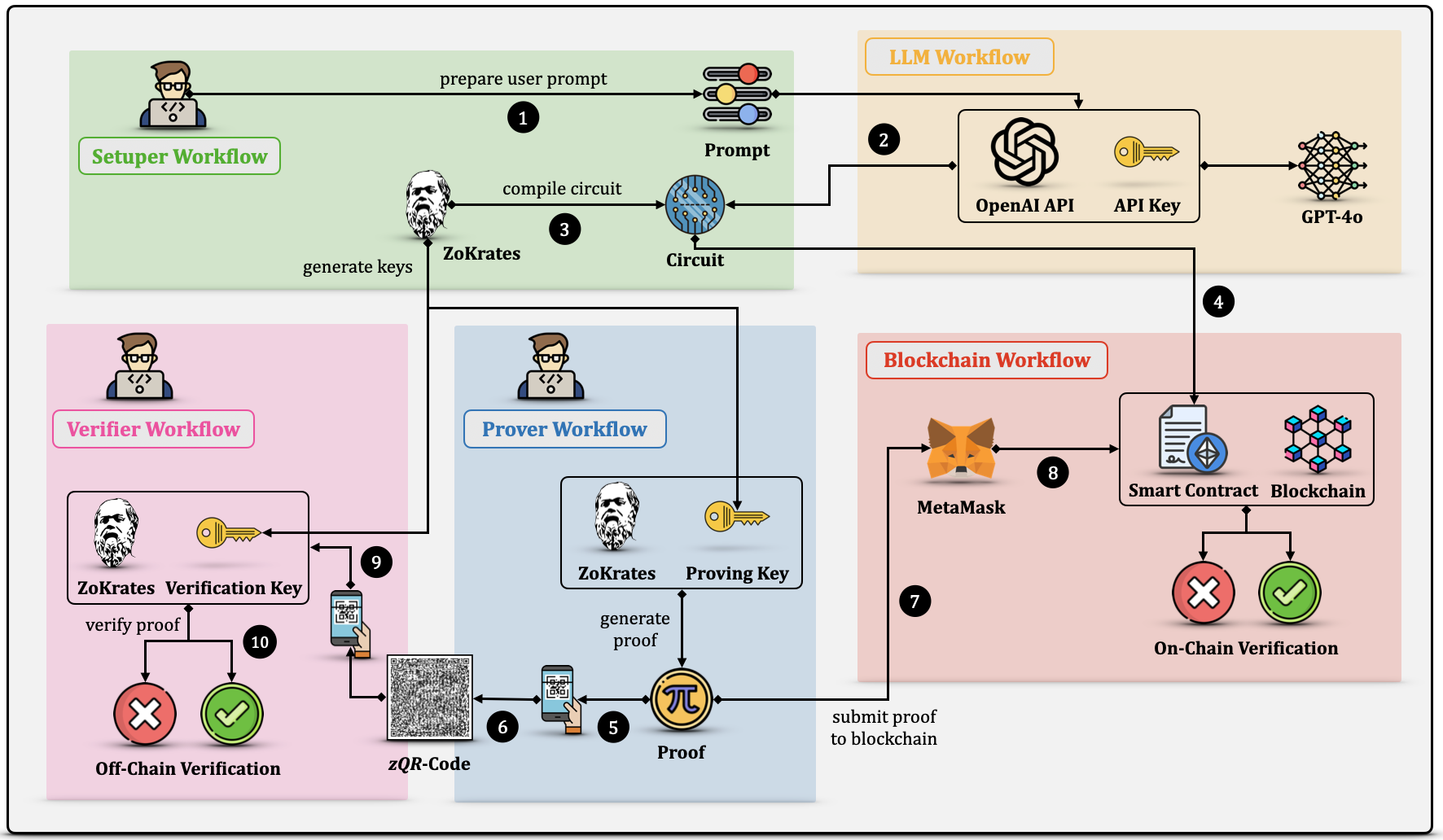}}
\caption{\textit{zQR} Framework: Issuer, Prover, Verifier, LLM and 
Blockchain Workflows}
\label{fig:architecture}
\end{figure*}

The \textit{zQR} framework is integrated with connections between roles as: issuer-LLM, issuer-blockchain, issuer-prover, issuer-verifier, prover-verifier and prover-blockchain. The overall architecture and connections are 
illustrated in Figure~\ref{fig:architecture}. In the \textit{zQR} framework, a heterogeneous adversarial environment is adopted where different roles operate under different threat assumptions. The role-based threat model is defined as follows:

\begin{itemize}
\itemsep0em
\item Issuers and verifiers are honest-but-curious, which strictly follow the framework rules but may also seek additional sensitive information beyond what is necessary.

\item Large-language models are also considered honest-but-curious where they never intentionally manipulate their responses about proof circuits. Nevertheless, their responses might be syntactically correct but semantically inconsistent with issuer intentions.

\item Provers are computationally bounded where they are not able to break standard zkSNARK security assumptions, but are considered potentially malicious that might attempt to deviate from the framework rules by manipulating inputs, constructing invalid proofs or delaying responsibilities to compromise framework correctness and security.
\end{itemize}

\subsection{The Issuer Workflow}

The issuer workflow covers the core responsibilities of the issuer: constructing the proof program; compiling it into an arithmetic circuit; performing trusted setup; and deploying the resulting smart contract to the blockchain. For high-level proof program construction (see Paths 1-2 in Figure~\ref{fig:architecture}), the issuer has two options as manual and automatic. In manual construction, the program instructions are written line-by-line by the issuer directly while in automatic construction the issuer prompts a large-language model to generate the program according to certain specifications, as described in the LLM workflow (see Section~\ref{subsection:llm-workflow}). In this context, the LLM acts as a semantic-to-arithmetic translator that accelerates and automates code generation. The manual approach is more time-consuming but permits finer control over the resulting circuit while the automatic approach reduces technical barriers. Once the proof program is finalized, the issuer compiles it into the corresponding arithmetic circuit through a zero-knowledge proof compiler and performs a trusted setup to generate the proving and verification keys (see Path 3 in Figure~\ref{fig:architecture}). The proving key is distributed to prover mobile platforms for proof generation while the verification key is distributed to verifier mobile platforms for proof verification. At this stage, the issuer is assumed to be honest and must irreversibly destroy the common reference string after key generation. Finally, the issuer deploys the resulting smart contract to the blockchain (see Path 4 in Figure~\ref{fig:architecture}), as described in the blockchain workflow (see Section~\ref{subsection:blockchain-workflow}).

\subsection{The Prover Workflow}

\begin{figure*}[htbp]
    \centering
    \begin{subfigure}[b]{0.42\textwidth}
        \includegraphics[width=\textwidth]{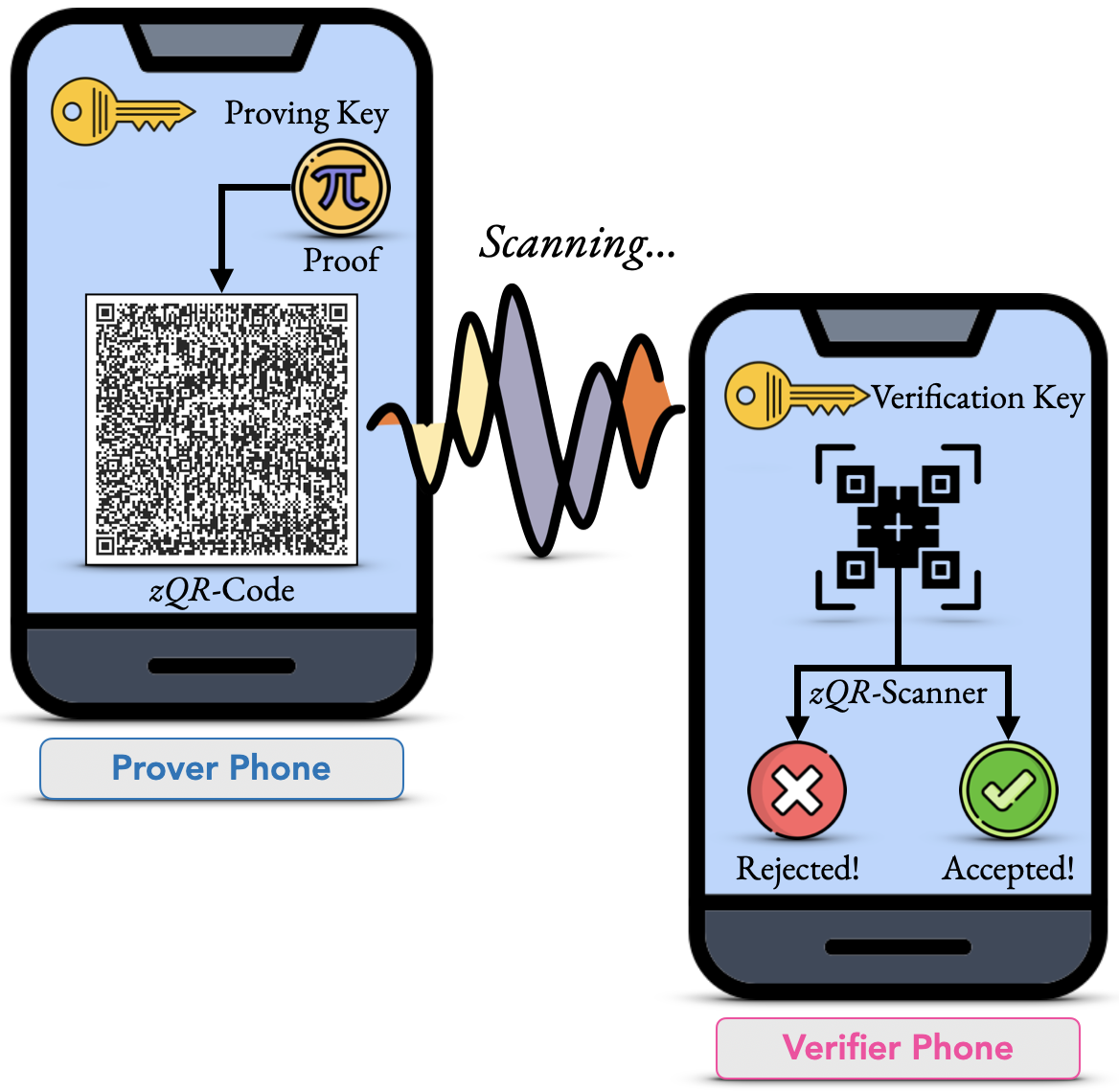}
        \caption{Proof Generation and Verification}
        \label{fig:phones}
    \end{subfigure}
    \begin{subfigure}[b]{0.57\textwidth}
        \includegraphics[width=\textwidth]{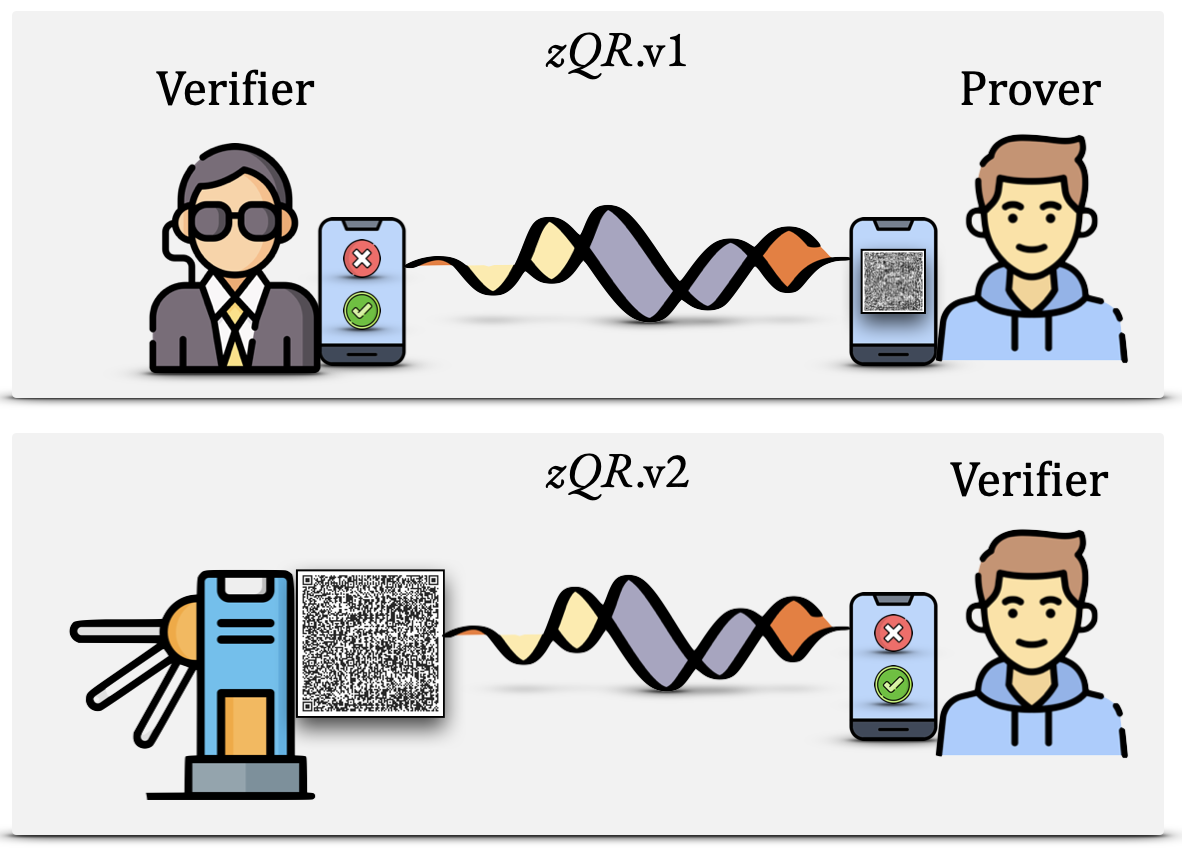}
        \caption{Comparison of \textit{zQR.v1} and \textit{zQR.v2}}
        \label{fig:vs}
    \end{subfigure}
    \caption{Scanning and Verifying Verifiable QR Codes in \textit{zQR.v1} and \textit{zQR.v2}}
    \label{fig:phones-vs}
\end{figure*}

The prover workflow covers the core responsibilities of the prover: generating a zero-knowledge proof; encoding it into a verifiable QR code; and submitting it to the blockchain for on-chain verification. For proof generation, the mobile interface first collects the necessary public and private inputs from the prover. The proof generator available in the interface locally generates a witness from the inputs and subsequently produces a succinct zero-knowledge proof (see Path 5 in Figure~\ref{fig:architecture}). The proof is then passed to the QR code generator, which transforms it into a verifiable QR code and renders it on screen for scanning by the verifier (see Figure~\ref{fig:phones}). Finally, the prover submits the public inputs and the proof to the blockchain for on-chain verification (see Paths 7-8 in Figure~\ref{fig:architecture}), as described in the blockchain workflow in Section~\ref{subsection:blockchain-workflow}. Several critical points must be noted regarding this workflow. First, the private inputs of the prover must be exposed only to local processing and must never leave the device. Second, the available mobile resources must be sufficient to handle the computational and memory load of proof generation.

\subsection{The Verifier Workflow}

The verifier workflow covers the core responsibilities of the verifier: decoding the verifiable QR code; and verifying the resulting proof off-chain. The verifier begins by scanning the verifiable QR code on the prover mobile device (see Figure~\ref{fig:phones}) and decodes it to recover the corresponding zero-knowledge proof (see Path 9 in Figure~\ref{fig:architecture}). The proof verifier available in the verifier interface verifies the recovered proof with the verification key (see Path 10 in Figure~\ref{fig:architecture}). Two practical points must be noted regarding this workflow. First, the verifier device must have sufficient memory to store the verification key. However, since verification key sizes remain constant and substantially smaller than proving keys across all circuit configurations, this requirement does not pose a significant task on mobile devices. Second, proof verification is computationally cheaper than proof generation and exhibits nearly constant temporal cost regardless of 
circuit complexity, which is suited for resource-constrained mobile environments. These interactions among the issuer, prover and verifier are the first variant of the framework, namely \textit{zQR.v1}.

\subsubsection{A Lightweight Verification Variant: \textit{zQR.v2}}

A more lightweight but restricted variant of the framework, namely \textit{zQR.v2}, can be derived from the same architecture, but by eliminating the prover workflow entirely. In this variant (see Figure~\ref{fig:vs}), the issuer first performs its standard responsibilities and then generates a verifiable QR code for a proof based on a proving key and a set of chosen inputs, thereby implicitly undertaking the prover responsibilities. The verifier is then expected to possess the correct verification key matching that proving key in order to verify the QR code successfully. This variant offers a verification key-based access control where only verifiers holding the correct key are authorized to verify. The security of the access control relies entirely on the confidentiality of the verification key, which means that a single compromised key breaks the access control for all parties. Refer to Algorithm~1 and Algorithm~2 for the pseudo-codes of the \textit{zQR.v1} and \textit{zQR.v2} variants, respectively.

\begin{table*}[htbp]
\small
\begin{center}
\begin{tabular}{|l|}
\hline
\cellcolor{gray!50} \textbf{Algorithm 1: \textit{zQR.v1} Framework}  \\
\hline
\textbf{1.} \textbf{Setup:} \\
\quad \textbf{1.1} Issuer manually constructs the proof circuit, $\mathsf{C}$. \\
\quad \textbf{1.2} \textbf{Prompt-LLM:} \\
\quad \quad \textbf{1.2.1} Issuer prompts LLM to generate circuit, $\mathsf{C \gets \mathcal{S}.LLM(u, x)}$. \\
\quad \textbf{1.3.} Issuer performs trusted setup, $\mathsf{(pk, vk) \gets \mathcal{S}.ZKSetup(1^\lambda, C)}$. \\
\quad \textbf{1.4.} \textbf{Deploy-OnChain:} \\
\quad \quad \textbf{1.4.1} Issuer deploys verification key to blockchain, $\mathsf{SC(addr) \gets \mathcal{S}.Deploy(vk)}$. \\
\textbf{2.} \textbf{Generate:} \\
\quad \textbf{2.1.} Prover accesses proving key $\mathsf{pk}$ and prepares witness $\mathsf{w}$ and public inputs $\mathsf{x}$. \\
\quad \textbf{2.2.} Prover generates a proof, $\mathsf{\pi \gets \mathcal{P}.ZKProve(pk, w, x)}$. \\
\quad \textbf{2.3.} Prover encodes proof to QR code, $\mathsf{qr^\pi \gets QREncode(\pi)}$. \\
\quad \textbf{2.4.} \textbf{Verify-OnChain:} \\
\quad \quad \textbf{2.4.1} Prover submits proof to contract for verification, $\mathsf{b \gets \mathcal{V}.ZKVerify(addr, \pi, x)}$. \\
\textbf{3.} \textbf{Verify:} \\
\quad \textbf{3.1.} Verifier accesses QR code and decodes it, $\mathsf{\pi \gets QRDecode(qr^\pi)}$. \\
\quad \textbf{3.2.} Verifier verifies the proof, $\mathsf{b \gets \mathcal{V}.ZKVerify(vk, \pi, x)}$. \\
\hline
\end{tabular}
\label{alg:zqrv1}
\end{center}
\small
\end{table*}

\begin{table*}[htbp]
\small
\begin{center}
\begin{tabular}{|l|}
\hline
\cellcolor{gray!50} \textbf{Algorithm 2: \textit{zQR.v2} Framework}  \\
\hline
\textbf{1.} \textbf{Setup:} \\
\quad \textbf{1.1} Issuer manually constructs the proof circuit, $\mathsf{C}$. \\
\quad \textbf{1.2} \textbf{Prompt-LLM:} \\
\quad \quad \textbf{1.2.1} Issuer prompts LLM to generate circuit, $\mathsf{C \gets \mathcal{S}.LLM(u, x)}$. \\
\quad \textbf{1.3.} Issuer performs trusted setup, $\mathsf{(pk, vk) \gets \mathcal{S}.ZKSetup(1^\lambda, C)}$. \\
\quad \textbf{1.4.} Issuer generates a verifiable proof, $\mathsf{\pi \gets \mathcal{S}.ZKProve(pk, w, x)}$. \\
\quad \textbf{1.5.} Issuer encodes proof to QR code, $\mathsf{qr^\pi \gets QREncode(\pi)}$. \\
\textbf{2.} \textbf{Verify:} \\
\quad \textbf{2.1.} Verifier accesses QR code and decodes it, $\mathsf{\pi \gets QRDecode(qr^\pi)}$. \\
\quad \textbf{2.2.} Verifier verifies the proof if the correct verification key is present, $\mathsf{b \gets \mathcal{V}.ZKVerify(vk, \pi, x)}$. \\
\hline
\end{tabular}
\label{alg:zqrv2}
\end{center}
\small
\end{table*}

\subsection{Integrating QR Codes}

\begin{figure*}[htbp]
\centerline{\includegraphics[width=0.99\textwidth]{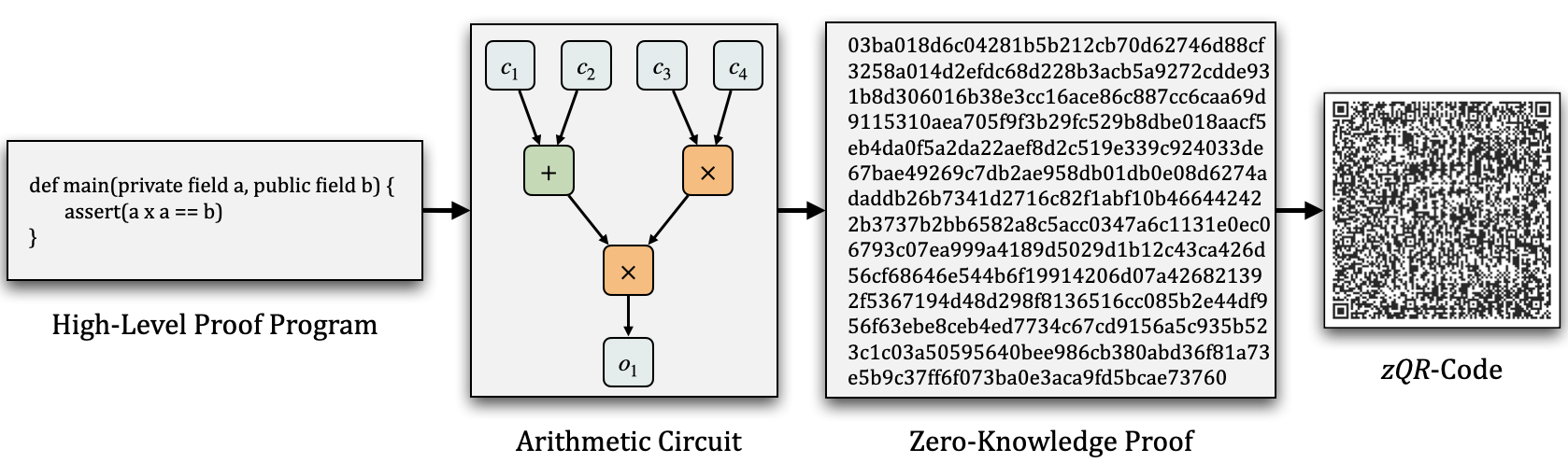}}
\caption{Transformation of High-Level Proof Program to Verifiable QR Code}
\label{fig:transformation}
\end{figure*}

The \textit{zQR} framework integrates verifiable QR codes for frictionless interaction between the prover and verifier 
workflows. By \textit{frictionless}, we refer to an interaction model that minimizes user effort, technical complexity and operational delay during proof generation and verification. This integration represents a transformation pipeline from a high-level proof program to an arithmetic circuit, from an arithmetic circuit to a zero-knowledge proof, and finally from a zero-knowledge proof to a 
verifiable QR code (see Figure~\ref{fig:transformation}). There exists a direct relationship between proof size and the symbol version of the QR code representing that proof. The larger the proof, the higher the symbol version required and consequently the larger and more visually complex the resulting QR code. Minimizing proof size is therefore an important consideration for producing more compact and user-friendly verifiable QR codes.

\subsection{The Large-Language Model (LLM) Workflow}
\label{subsection:llm-workflow}

The \textit{zQR} framework connects to the LLM workflow via asynchronous HTTP requests for automatic proof circuit 
generation from user prompt. This workflow is designed to lower the technical barriers of the framework by eliminating the need for issuers to have expertise in zero-knowledge proof. In this workflow, the issuer defines the statement 
to be proven by specifying the public and private inputs. This prompt is combined with a pre-defined system prompt containing instructions, boundaries and constraints. The issuer submits the resulting prompt to the external OpenAI API using a valid API key and waits for the response (see Path 2 in Figure~\ref{fig:architecture}). From a privacy perspective, communication with the external API does not require any sensitive witness data since the workflow operates only at the level of proof specifications. However, the circuit structure itself might reveal information about the nature of the application being built. The LLM-generated circuit is also not guaranteed to be semantically correct with respect to issuer intentions. Therefore, human-in-the-loop validation of the circuits is recommended before proceeding to the trusted setup phase. This workflow also introduces additional latency due to LLM response generation and network transmission, which should be taken into account for time-sensitive deployments. Refer to the work \cite{ismayilov2025} for LLM-based circuit generation.

\subsection{The Blockchain Workflow}
\label{subsection:blockchain-workflow}

The \textit{zQR} framework connects to the blockchain workflow via crypto-wallets (i.e., MetaMask \cite{metamask}) for smart contract deployment and on-chain proof verification. The contract deployment is performed by the 
issuer which compiles and deploys the contract to an EVM-compatible blockchain network (see Path 4 in Figure~\ref{fig:architecture}). Once deployed, the contract 
is publicly accessible at its address. It is the issuer that covers the cost of that one-time deployment. Later, verifiers can benefit from on-chain verification without incurring any additional costs. The on-chain proof verification is performed by the prover which externally calls the proof-verifying function (i.e., \textit{verifyTx}) in the contract by submitting the public inputs and the proof itself. This function is defined with the \textit{view} modifier, which means that it is read-only and does not modify the blockchain state and therefore is gas-free. However, if \textit{verifyTx} is invoked indirectly by a state-changing function (e.g.,  transfers assets upon successful proof verification), the prover would need to cover the corresponding gas costs. Note that any modification to an existing proof circuit requires its corresponding smart contract to be redeployed since each contract is tightly coupled to a specific verification key that is unique to the circuit it is generated from. This redeployment incurs an additional one-time financial cost and must be planned accordingly in production deployments.

\section{Security Threat Discussion of \textit{zQR} Framework}

In this section, we discuss the security threat landscape of the proposed framework from the zero-knowledge proof, QR code, blockchain and operational perspectives. The threat model assumes honest-but-curious issuers and verifiers; and computationally-bounded malicious provers. Stronger adversarial settings (including malicious issuers, malicious verifiers and side-channel attacks) are acknowledged as open challenges and are outside the scope of the current work.

\subsection{Zero-Knowledge Proof Threats}

\textit{Compromising toxic waste.} The trusted setup phase produces a common reference string (i.e., toxic waste) where compromising it violates protocol soundness by enabling adversarial parties to forge proofs for false statements. To mitigate this risk, a secure multi-party computation ceremony might be adopted, which distributes the setup responsibility across multiple nodes such that soundness holds as long as at least one node is honest in return of additional coordination complexity of these nodes. Under the current framework assumptions, issuers are modeled as honest and are expected to irreversibly destroy the toxic waste after key generation. Exploring trustless alternatives (e.g., zkSTARKs \cite{zkstarks}) is a promising future direction.

\textit{Proof replaying.} The proofs in the framework are non-interactive and publicly verifiable, which means that a copy of a valid proof instance can be repeatedly scanned and verified with the correct verification key. This enables an adversary to impersonate a legitimate prover without ever knowing the private witness. This attack might also be carried out through QR code cloning where a verifiable QR code is photographed and reproduced. A practical deployment may maintain an on-chain registry of proof identifiers to detect replayed proofs.

\subsection{QR Code Threats}

\textit{Over-the-shoulder attacks.} The prover mobile interface requires private inputs to be entered directly on screen since the framework does not persistently store these inputs anywhere. Adversaries might attempt to observe these values through shoulder surfing or visual surveillance. In addition, if a prover device is already compromised (i.e., via key-loggers), private inputs might be exposed or manipulated before they reach the proof generation. Unfortunately, this threat cannot be fully eliminated at the framework level.

\textit{QR code payload injection.} Maliciously-crafted QR codes might inject malformed proof payloads with the goal of exploiting vulnerabilities in QR decoding libraries. To reduce this risk, strict proof format validation should be enforced in the verifier workflow prior to any verification. Our framework relies on a popular QR code library for encoding and decoding proofs.

\subsection{Blockchain Threats}

\textit{Transaction replaying.} Rather than replaying the proof itself, adversaries might replay proof-carrying blockchain transactions to repeatedly trigger verification of the same proof. Introducing unique nonce identifiers per transaction might mitigate this within the same blockchain network. However, front-running still remains possible where an adversary actively monitors the pending transactions and resubmits them with higher gas fees to commit earlier than the original transaction owner.

\textit{Smart contract attacks.} ZoKrates \cite{eberhardt2018} automatically generates template-based smart contracts that are bound to their corresponding circuits. These contracts might be susceptible to well-known smart contract vulnerabilities (e.g., re-entrancy, integer underflow and overflow). The security of the blockchain workflow is therefore as strong as the underlying guarantees of ZoKrates. Reinforcing the cryptographic security of these contracts is outside the scope of this work. 

\subsection{Operational Threats}

\textit{Denial-of-service.} Adversaries might attempt to stall the framework by flooding with a large number of requests by congesting blockchain interactions or overwhelming the LLM service endpoint. Mitigation strategies might include rate limiting, resource monitoring and early detection alarms. Evaluating the framework performance under growing loads is also identified as a future work.

\textit{Prompt manipulation.} The LLM workflow accepts natural language specifications from issuers to automatically generate proof circuits. An adversary with access to the issuer interface might craft malicious prompts to produce flawed or weakened circuits that can pass syntactic compilation but fail to enforce semantics. Mitigation might include human-in-the-loop validation of the circuits before trusted setup is performed.

\section{Experimental Study}

The experiments are conducted on three different hardware platforms. \textit{zQR-macOS} is a MacBook Air with an Apple M2 chip, 8 GB of memory and 8 CPU cores. \textit{zQR-iOS} is an iPhone 15 with an Apple A16 Bionic chip, 6 GB of memory and a 6-core CPU. \textit{zQR-aOS} is a Samsung A2 with a quad-core processor, 2 GB of memory and 4 CPU cores. These three platforms represent a range of hardware capabilities from a high-performance desktop-class machine to a severely resource-constrained Android device. The user interfaces for the issuer, prover and verifier workflows rely on the 
\textit{zokrates.js} \cite{eberhardt2018} library for proof generation and verification and the \textit{qrcode.js} library for QR code encoding and decoding. The LLM workflow uses OpenAI GPT-4.1 \cite{gpt4o} as the large-language 
model, which is connected to the interfaces via HTTP requests and responses. The blockchain workflow uses the Solidity language deploys the contract to the Ethereum Sepolia 
network. The smart contract is bridged to the interfaces via the MetaMask wallet \cite{metamask}. The characteristics of the proof circuits used across our experiments are given in Table~\ref{table:circuits}.

\begin{table*}[htbp]
\caption{Zero-Knowledge Proof Circuit Specifications}
\footnotesize
\begin{center}
\begin{tabular}{lrrrrrr}
\hline
\cellcolor{gray!50} \textbf{Circuit} & 
\cellcolor{gray!50} \textbf{Private Inputs} & 
\cellcolor{gray!50} \textbf{Public Inputs} & 
\cellcolor{gray!50} \textbf{Constraints} & 
\cellcolor{gray!50} \textbf{Proving Key} & 
\cellcolor{gray!50} \textbf{Verification Key} & 
\cellcolor{gray!50} \textbf{Proof Size} \\
\hline
C1 & 1 & 1 & 2 & 2 KB & 2 KB & 1 KB \\
C2 & 1 & 2 & 3,309 & 1.3 MB & 2 KB & 1 KB \\
C3 & 2 & 4 & 48,887 & 19.7 MB & 2 KB & 1 KB \\
C4 & 3 & 6 & 52,196 & 20.8 MB & 3 KB & 1 KB \\
C5 & 8 & 4 & 97,773 & 39.5 MB & 2 KB & 1 KB \\
C6 & 8 & 6 & 145,199 & 62.9 MB & 3 KB & 1 KB \\
C7 & 8 & 2 & 232,309 & 90.6 MB & 2 KB & 1 KB \\
C8 & 8 & 2 & 330,077 & 138.5 MB & 2 KB & 1 KB \\
\hline
\multicolumn{4}{l}{\footnotesize KB: Kilobytes, MB: Megabytes} \\
\end{tabular}
\label{table:circuits}
\end{center}
\end{table*}

\subsection{Temporal Cost of Proof Generation and Verification on Mobile Platforms}

\begin{figure*}[htbp]
    \centering
    \begin{subfigure}[b]{0.49\textwidth}
        \includegraphics[width=\textwidth]{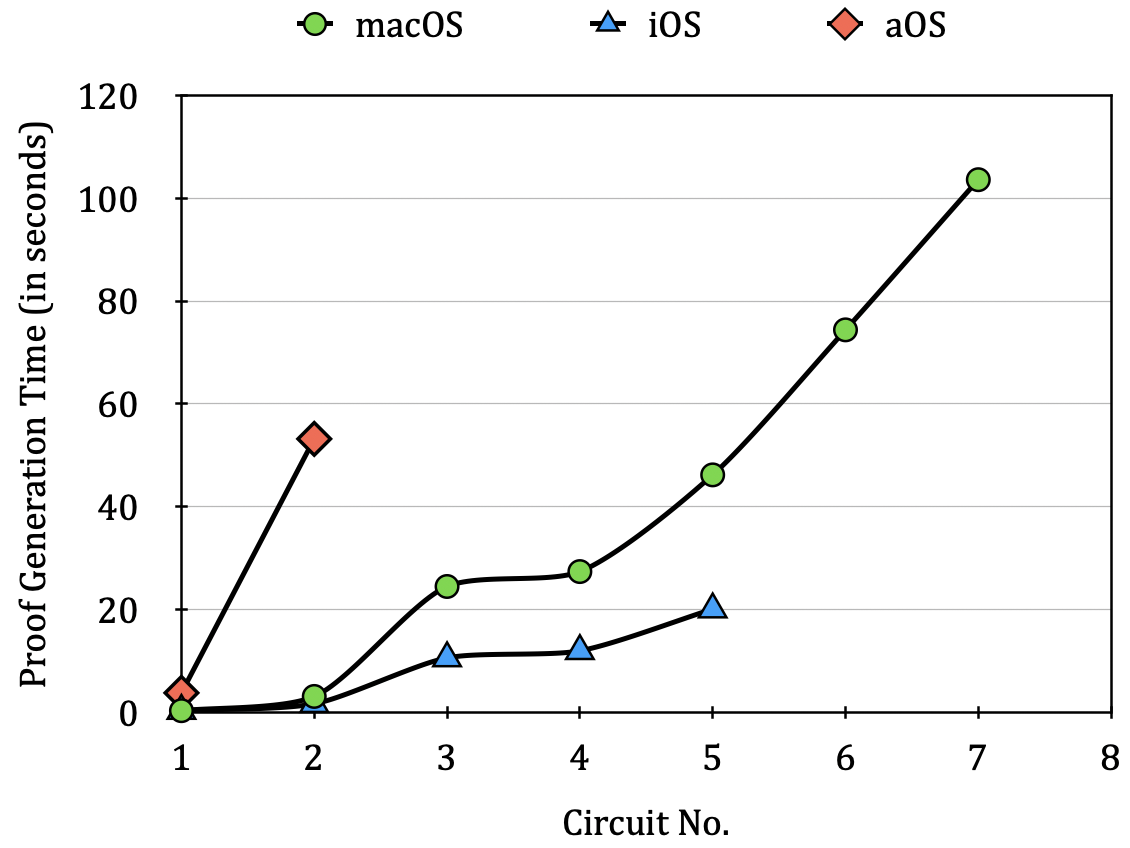}
        \caption{Proof Generation Time}
        \label{fig:generation-time}
    \end{subfigure}
    \begin{subfigure}[b]{0.47\textwidth}
        \includegraphics[width=\textwidth]{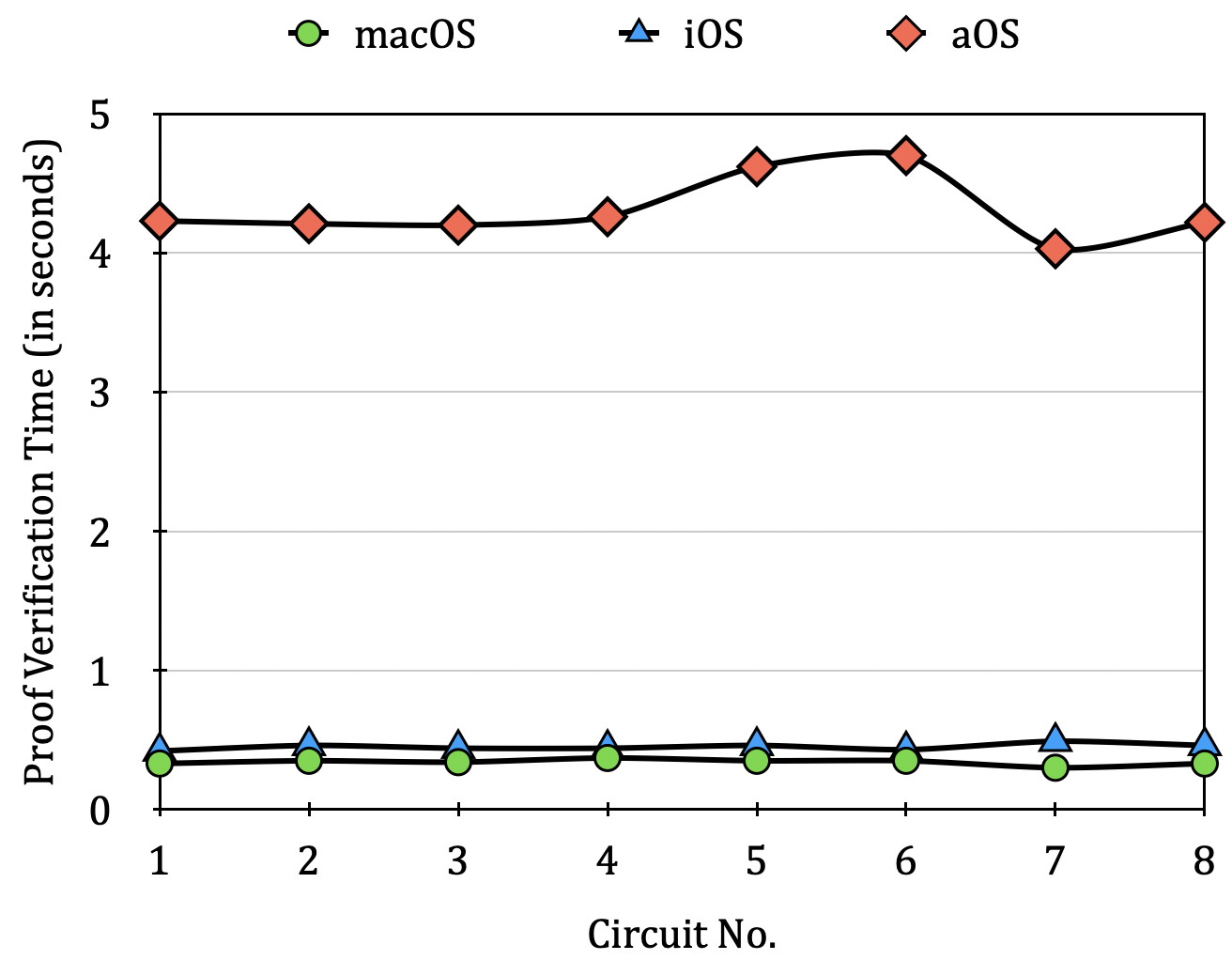}
        \caption{Proof Verification Time}
        \label{fig:verification-time}
    \end{subfigure}
    \begin{subfigure}[b]{0.49\textwidth}
        \includegraphics[width=\textwidth]{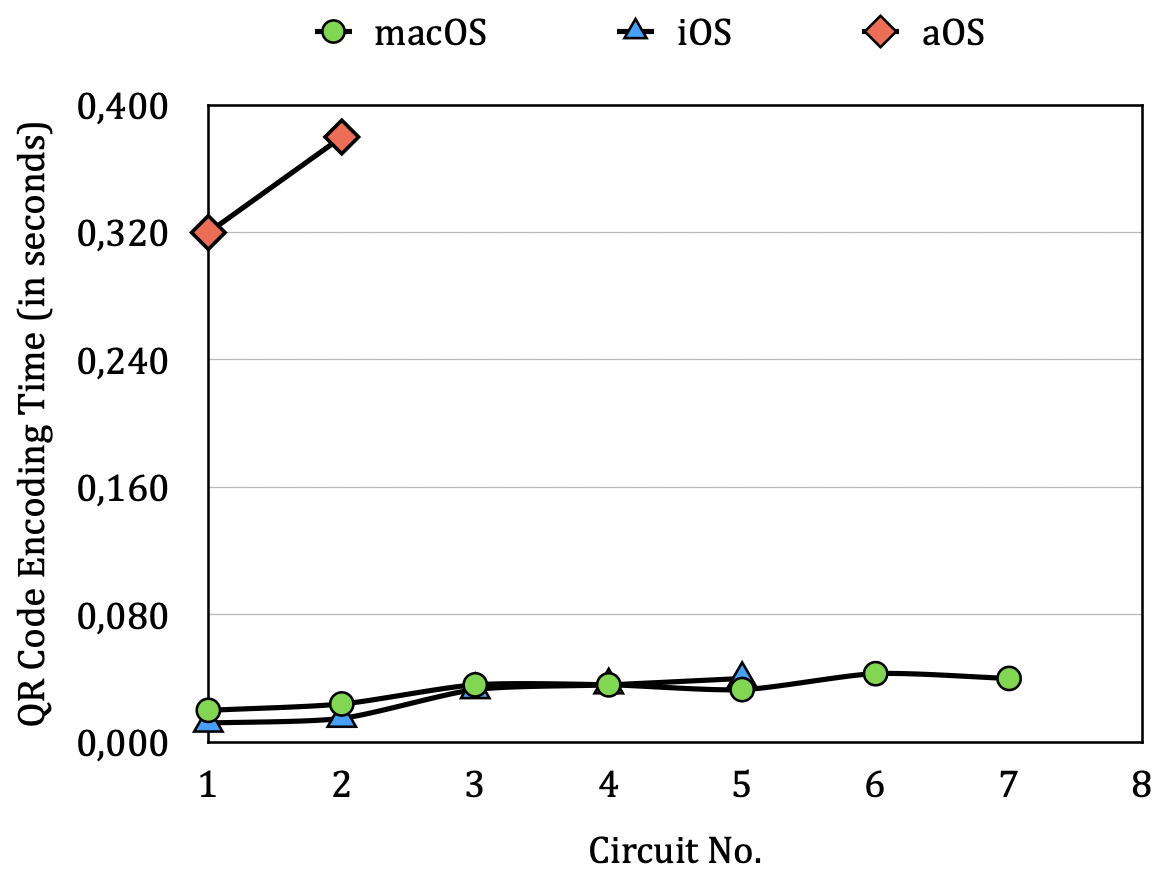}
        \caption{QR Code Encoding Time}
        \label{fig:encoding-time}
    \end{subfigure}
    \begin{subfigure}[b]{0.48\textwidth}
        \includegraphics[width=\textwidth]{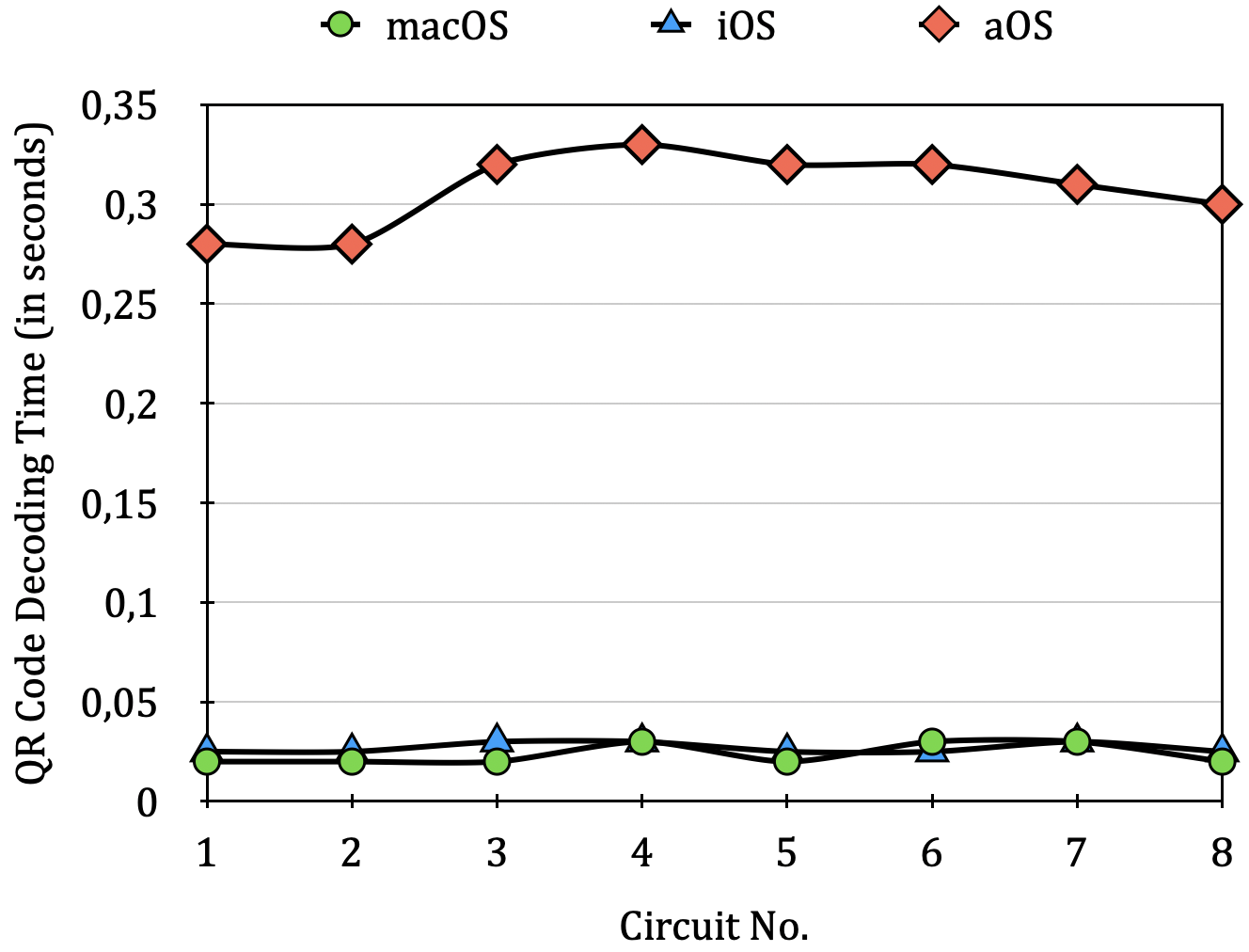}
        \caption{QR Code Decoding Time}
        \label{fig:decoding-time}
    \end{subfigure}
    \caption{Temporal Costs of Proof Generation/Verification and QR Code Encoding/Decoding Operations}
    \label{fig:temporal}
\end{figure*}

Proof generation and verification costs are specifically critical for resource-constrained mobile platforms. This experiment measures these costs for the circuits given in 
Table~\ref{table:circuits} and presents the temporal results in Figure~\ref{fig:temporal} for three devices of \textit{macOS}, \textit{iOS} and \textit{aOS}. 

\textit{Observation 1. Proof generation time grows with increasing circuit complexity.} Figure~\ref{fig:generation-time} shows a positive correlation between circuit complexity and proof generation time where the computational effort grows proportionally with the number of arithmetic constraints. For instance, the first circuit with 2 constraints requires approximately 0.35 seconds to generate a proof on \textit{macOS} whereas the seventh circuit with 232,309 constraints requires approximately 103.6 seconds on the same device. Among all devices, \textit{aOS} exhibits substantially higher proving times even for smaller circuits. Furthermore, \textit{aOS}, \textit{iOS} and \textit{macOS} are unable to generate proofs for the 3rd, 6th and 8th circuits respectively due to memory limitations on each device. On \textit{iOS}, proofs can be generated for circuits up to approximately 100,000 constraints within a practically acceptable time of around 20 seconds. Overall, Figure~\ref{fig:generation-time} confirms that proof 
generation is the primary computational bottleneck of the framework. Several mitigation strategies exist in the 
literature to address this bottleneck. Vertical batching \cite{gu2024} divides the proof program into several sub-programs so that proofs are generated sequentially, reducing peak computational load. Choosing zkSNARK-friendly hash functions \cite{kondyrev2024} (e.g., Poseidon \cite{grassi2021}) drastically reduces circuit size. Outsourcing proof generation to external services via API calls \cite{ismayilov2025} offloads the computation entirely from the mobile device.

\textit{Observation 2. Off-chain proof verification time remains constant regardless of circuit complexity.} Figure~\ref{fig:verification-time} shows that off-chain proof verification is cheaper than proof generation and exhibits nearly constant temporal cost across circuits of varying sizes. Verification latency is also more predictable and stable across devices within a narrow temporal range. On 
\textit{macOS} and \textit{iOS}, verification time fluctuates slightly between 0.3 and 0.5 seconds with minimal sensitivity to circuit complexity. \textit{aOS} exhibits relatively higher verification latency of approximately 4.2 to 4.7 seconds, but this is still within an acceptable range for practical verification scenarios. These results further indicate that the lightweight \textit{zQR.v2} variant is more suitable for resource-constrained mobile platforms since it eliminates proof generation and retains only the constant-time verification operation. 

\subsubsection{Temporal Cost of QR Code Encoding and Decoding on Mobile Platforms}

The central claim of this work is based on the frictionless representation of zero-knowledge proofs via verifiable QR codes. From this perspective, measuring the time to encode and decode is critical to assess the practical feasibility of the framework.

\textit{Observation 3. QR code encoding and decoding times remain constant regardless of circuit complexity.} Figure~\ref{fig:encoding-time} and Figure~\ref{fig:decoding-time} indicate that both encoding and decoding operations exhibit approximately constant and stable temporal cost across all circuits. On \textit{macOS} and \textit{iOS}, encoding time remains below 0.05 seconds even for the largest circuits, while \textit{aOS} shows a higher but still stable encoding time of approximately 0.32 to 0.38 seconds. This stability is a direct consequence of the constant proof size of the zkSNARK protocol regardless of circuit complexity (see Table~\ref{table:circuits}). Devices that are unable to generate proofs for certain circuits due to memory limitations do not naturally perform encoding operations. Note that potential environmental factors (e.g., lighting, camera focus and motion blur) are explicitly excluded from 
these measurements.

\textit{Observation 4. QR code encoding and decoding operations are feasible on all mobile platforms.} QR code decoding time is measured from the moment the device camera successfully detects and recognizes the QR code pattern. Figure~\ref{fig:decoding-time} shows that decoding times are consistently low and nearly constant across all circuits. 
On \textit{macOS} and \textit{iOS}, decoding time remains within approximately 0.02 to 0.03 seconds across all circuits. \textit{aOS} again exhibits higher values of approximately 0.28 to 0.33 seconds, but these remain stable regardless of circuit complexity. Even on the most resource-constrained device evaluated, the combined encoding and decoding overhead introduced by the QR code layer remains below one second. 

\textit{Observation 5. QR code symbol version varies with error correction level for a fixed proof size.} As shown in Table~\ref{table:circuits}, the proof size remains constant for zkSNARKs. Specifically, encoding a verifiable proof requires symbol version 19 for the low error correction level, symbol version 22 for the medium level, symbol version 27 for the quartile level and symbol version 32 for the high level. These results reflect the trade-off in QR code 
design where higher error correction levels reserve a larger space for redundancy, thereby reducing the available capacity for the actual payload. From a practical standpoint, the low level produces the most compact verifiable QR code and is 
recommended for screen-to-camera scanning scenarios. Higher levels are recommended when the verifiable QR code is printed on a physical surface that might be subject to damage.

\subsection{Financial Cost of Smart Contract Executions on Blockchain}
\label{section:financial-cost}

\begin{figure*}[htbp]
    \centering
    \begin{subfigure}[b]{0.42\textwidth}
        \includegraphics[width=\textwidth]{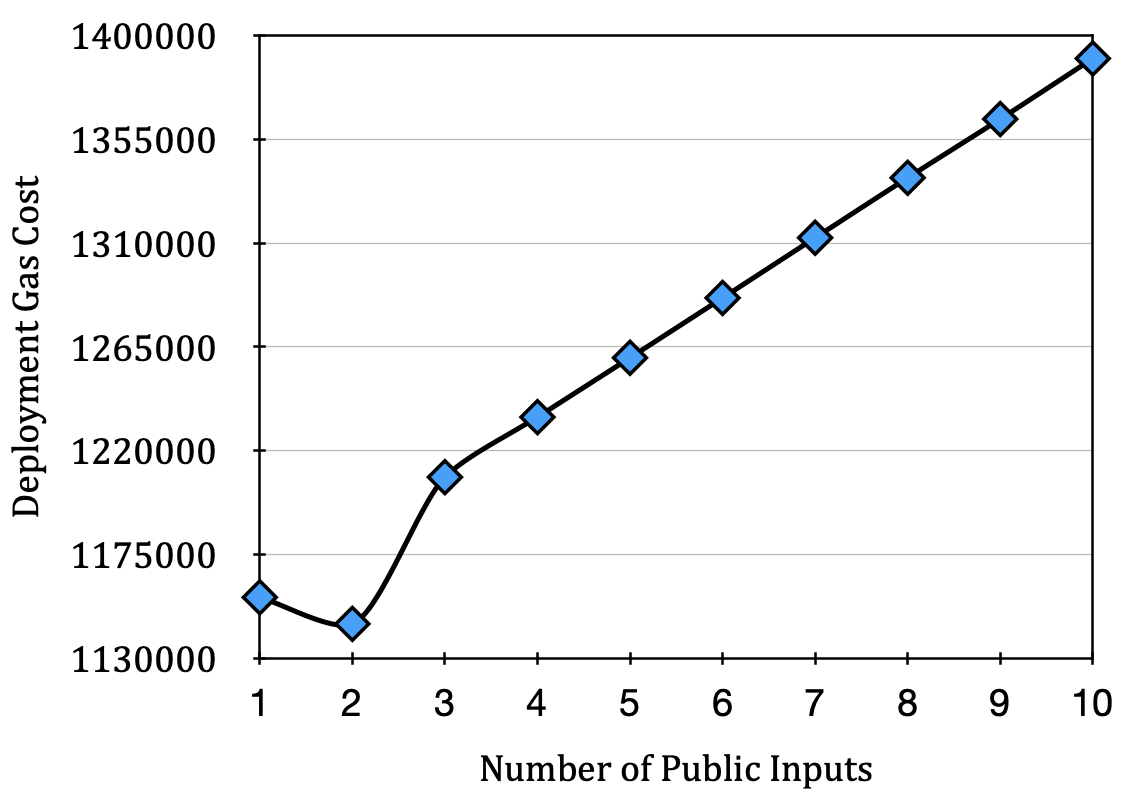}
        \caption{In Ethereum Gas Units}
        \label{fig:in-gas}
    \end{subfigure}
    \begin{subfigure}[b]{0.40\textwidth}
        \includegraphics[width=\textwidth]{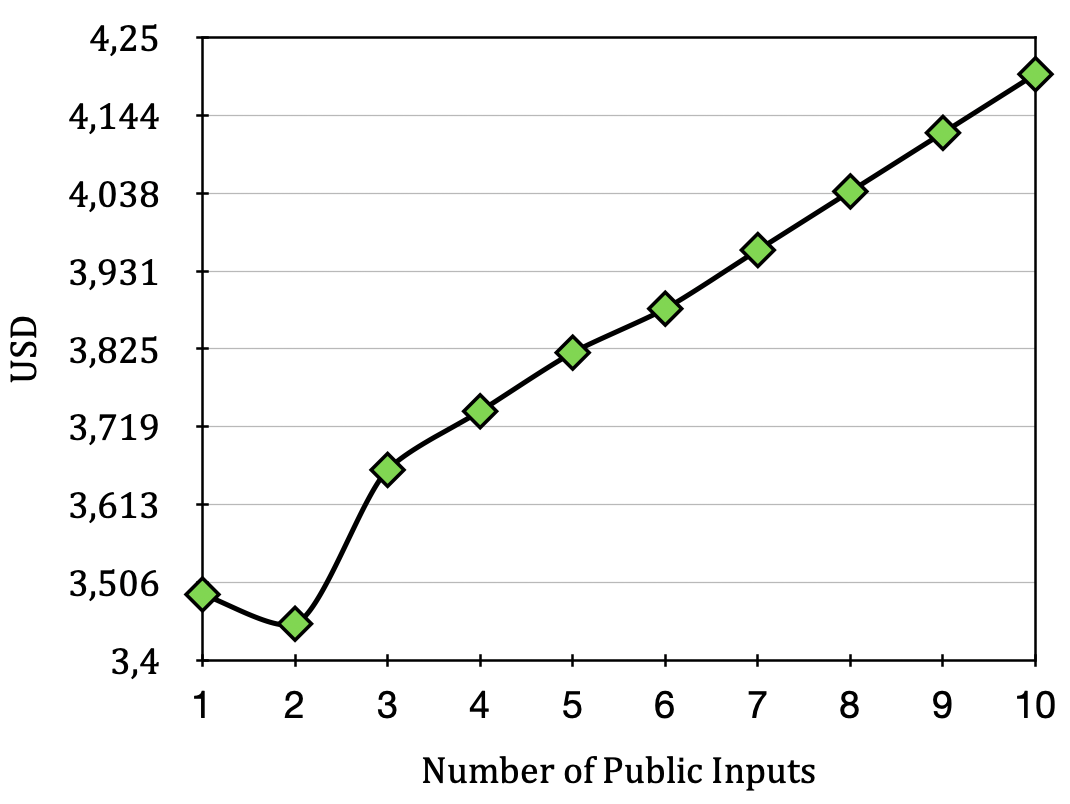}
        \caption{In American Dollars}
        \label{fig:in-usd}
    \end{subfigure}
    \caption{Financial Cost of Smart Contract Deployments over Varying Number of Public Inputs}
    \label{fig:gas-cost}
\end{figure*}

This experiment measures the financial cost of deploying a single proof-verifying smart contract to the blockchain and presents the results in Figure~\ref{fig:gas-cost} across varying numbers of public inputs. These costs may vary across different blockchain platforms, EVM versions and ZoKrates versions. The exchange rate used for financial conversion is 1.5 Gwei per gas unit and 2,023 USD per ETH.

\textit{Observation 6. Smart contract deployment cost grows proportionally with the number of public inputs.} On-chain proof verification in ZoKrates requires only the public inputs and the proof. Since the proof size always remains constant regardless of circuit complexity, the primary factor in contract deployment gas consumption is the number of public inputs. As clearly visible in Figure~\ref{fig:gas-cost}, gas consumption increases linearly with each additional public input at an approximate rate of 30,000 gas units per input. Converting these measurements into financial costs, the minimum deployment cost with two public inputs is approximately 0.00171 ETH (approx. 3.45 USD) while the maximum cost for ten public inputs reaches 0.00208 ETH (approx. 4.20 USD). 

\textit{Observation 7. On-chain proof verification does not incur direct costs for provers.} ZoKrates implements the verification function as follows:
\begin{align}
\mathsf{function~verifyTx(Proof~memory~proof,~uint[n]~memory~input)} \mathsf{public~view~returns~(bool~r)}
\end{align}
\noindent
with the \textit{view} modifier, which means that the function only reads blockchain state data without writing to it. However, it is important to note that if \textit{verifyTx} is invoked indirectly by a state-changing function, the associated gas costs are required to be covered.

\subsection{Circuit Complexity of Zero-Knowledge Proofs}
\label{section:circuit-complexity}

This experiment focuses on the complexity characteristics of the eight proof circuits in Table~\ref{table:circuits} in terms of the number of private inputs, public inputs, arithmetic constraints, proving key size, verification key size and proof size.

\textit{Observation 8. Circuit constraint count and proving key size grow proportionally with computational complexity.} Cryptographic operations such as hashing and public-key encryption in ZoKrates contribute to arithmetic circuit expansion. The first and simplest circuit C1 proves that the square of a private value equals a public value and requires no cryptographic operations, producing only 2 constraints and 
a proving key of 2 KB. In contrast, the last and most complex circuit C8 proves that the public root hash of a Merkle tree with 8 leaf values is correctly computed through 7 SHA-256 hashing operations, producing a total of 330,077 constraints and a proving key of 138.5 MB. The management of proving keys becomes particularly critical in resource-constrained environments since excessively large proving keys may exceed available device memory and cause proof generation to fail.

\textit{Observation 9. Verification key size and proof size remain constant across all circuits.} Despite the constraint count increasing from 2 to 330,077 across the circuits, the verification key size remains within a narrow range and the proof size remains fixed across all circuits. The constant verification key size is particularly important for the blockchain workflow since ZoKrates embeds the verification key directly into the smart contract. Any growth 
in verification key size would directly increase contract deployment gas costs. Similarly, the constant proof size  presents predictable and uniform QR code encoding requirements across all circuits, simplifying symbol version selection. It ensures that the same QR configuration can be used regardless of the circuit. 

\section{Further Discussion on Practical Applications}

In this section, we discuss the potential applications of the \textit{zQR} framework across several privacy-sensitive domains. These applications are given to show the extent of the framework applicability:

\begin{itemize}
\itemsep0em

\item \textit{Physical Access Control.} Most of the current access control systems (e.g., biometric scanners) rely on explicitly verifying identity attributes, which requires the disclosure of personal information. The \textit{zQR} framework enables users to prove their access credentials through verifiable QR codes without revealing their actual identities. It is also applicable to traditional attribute-based verification where users prove the statements (e.g., being over 18 years of age, being physically present within campus) without disclosing the attributes themselves.

\item \textit{Payment and Purchase.} Current payment systems collect sensitive transaction details and record transaction histories in centralized databases for auditing (e.g., tax calculations). However, these records enable financial profiling and behavior analysis of customers. The \textit{zQR} framework has the potential to allow anonymous purchasing through a simple QR code interaction. Zero-knowledge proof also enables tax calculations to be carried out privately without disclosing the financial details.

\item \textit{Anonymous Membership and Subscription.} Existing membership systems link subscriptions to real identities, enabling service providers to track members and their behaviors over time. The \textit{zQR} framework might allow members to prove that their memberships are active without revealing further personal details. This application extends naturally to anonymous event ticketing where ticket holders prove validity without disclosing their identities.

\item \textit{Anonymous Voting.} A substantial body of work in the literature proposes cryptographically strong voting protocols while overlooking their real-world adoption \cite{panja2018, wu2023, miao2023}. The \textit{zQR} framework is positioned to bridge this gap by allowing polling stations to challenge voters to anonymously prove their eligibility and non-participation through verifiable QR codes.

\item \textit{Border Control and Anonymous Traveling.} Current border control procedures require travelers to disclose passports, biometric data, travel history and visa information for multiple jurisdictions and decision-making authorities. Within the \textit{zQR} framework, travelers can anonymously prove their compliance through a simple QR code scanning interaction. This represents a promising direction for the future of privacy-preserving travel.

\end{itemize}

\section{Further Discussion on Current Limitations and Future Directions}

In this section, we discuss the current limitations of the \textit{zQR} framework and identify corresponding future research directions.

\begin{itemize}
\itemsep0em
    
\item \textit{Trusted Issuer Workflow.} The zkSNARK implementation relies on a common reference string to generate keys, which introduces the trust assumption that soundness is compromised if the string is not properly destroyed after key generation. ZoKrates itself partially addresses this by proposing a multi-party ceremony to distribute the setup responsibilities across multiple nodes \cite{eberhardt2018}. Exploring trusted-setup-free alternatives (e.g., zkSTARKs \cite{zkstarks}) is also a promising future direction at the cost of larger proof sizes, which might exceed the information capacity of QR code symbol versions.

\item \textit{Stronger Threat Model.} The current proof-of-concept implementation assumes honest-but-curious roles for issuers and verifiers. However, real-world scenarios might require stronger adversarial assumptions where malicious issuers might compromise the trusted setup ceremony or malicious verifiers may selectively reject correct proofs or introduce artificial verification delays. Furthermore, side-channel attacks on mobile devices (e.g., timing attacks during proof generation) have not been considered in the current threat model. 

\item \textit{Proof Size and Symbol Versions.} The zkSNARK protocol produces fixed-size and succinct proofs, which makes their QR code representation feasible within symbol version 19. However, minimizing the symbol version is still a desirable objective for producing more compact and user-friendly verifiable QR codes. Future work may explore alternative proof serialization formats or more compact elliptic curve representations to reduce the payload size without compromising verification correctness.

\item \textit{Stronger Blockchain Integration.} The blockchain workflow of the \textit{zQR} framework is currently tailored to EVM-compatible blockchains. Extending verification and state synchronization to non-EVM ecosystems and multi-chain environments is challenging in terms of gas consumption, available proof verification libraries and cross-chain bridging. However, it would significantly broaden the applicability of the framework.

\item \textit{Better LLM Performance.} Although LLM performance is not the primary focus of this work, faster and more semantically accurate models would directly contribute to the reliability of automatic circuit generation. Employing stronger models with greater reasoning capacity, specifically tailored for proof circuit generation, might be promising. The following work \cite{ismayilov2025} empirically shows the feasibility of LLM-based circuit generation up to approximately 200,000 circuit constraints.

\item \textit{Scalability.} A real-world deployment of the \textit{zQR} framework raises scalability concerns across multiple dimensions (e.g., the number of individual deployments, the number of blockchain accounts in on-chain verification, the throughput of the blockchain network, the latency to external LLM services). Addressing these scalability dimensions is necessary to a production-level deployment of the framework.

\end{itemize}

\section{Conclusion}

In this work, we focus on the real-world adoption of zero-knowledge proof and proposed a novel proof-of-concept QR-driven zkSNARK proof verification, specifically targeting mobile platforms. The framework operates through the interconnections of three core workflows for the issuer, prover and verifier; and two optional integrations for large-language models and blockchain. The framework includes two variants: \textit{zQR.v1} for full three-party 
interaction and \textit{zQR.v2} as a lightweight alternative where the issuer undertakes the prover responsibilities. We discuss the potential threats by identifying risks across multiple attack surfaces including zero-knowledge proof attacks, QR code attacks and blockchain-level attacks. We acknowledge that stronger adversarial settings (e.g., malicious issuers, side-channel attacks) remain open challenges for future work. We also carry out an evaluation to measure performance in terms of temporal and financial costs. The results reveal proof generation remains the primary computational bottleneck and may take up to several minutes depending on circuit complexity. On-chain proof verification is gas-free for provers while deployment costs remain within an acceptable one-time range for issuers. These findings collectively demonstrate the feasibility of \textit{zQR} as a proof-of-concept framework. The potential use cases of the framework span a wide range of privacy-sensitive domains including physical access control, anonymous voting, anonymous traveling and privacy-preserving payment systems. Current limitations including the trusted setup requirement, single toolchain dependency and scalability constraints are clearly identified and a collection of future research directions are provided as well. To conclude, we believe that QR codes will be among the key enabling technologies for the broader popularization of zero-knowledge proof as privacy concerns continue to grow in modern societies. The \textit{zQR} framework represents a step in this direction, demonstrating that a more privacy-preserving, publicly-verifiable and auditable world is possible.

\bibliographystyle{ACM-Reference-Format}
\bibliography{sample-base}

@String{Computing = "Computing" }

@String{Computer = "{IEEE} Computer" }

@String{Springer = "Springer-Verlag" }

@CONFERENCE{eberhardt2018,
  title={Zokrates-scalable privacy-preserving off-chain computations},
  author={Eberhardt, Jacob and Tai, Stefan},
  booktitle={2018 IEEE International Conference on Internet of Things (iThings) and IEEE Green Computing and Communications (GreenCom) and IEEE Cyber, Physical and Social Computing (CPSCom) and IEEE Smart Data (SmartData)},
  pages={1084--1091},
  year={2018},
  organization={IEEE}
}

@ARTICLE{belles2022,
  title={Circom: A circuit description language for building zero-knowledge applications},
  author={Bell{\'e}s-Mu{\~n}oz, Marta and Isabel, Miguel and Mu{\~n}oz-Tapia, Jose Luis and Rubio, Albert and Baylina, Jordi},
  journal={IEEE Transactions on Dependable and Secure Computing},
  volume={20},
  number={6},
  pages={4733--4751},
  year={2022},
  publisher={IEEE}
}

@CONFERENCE{ismayilov2025,
  title={Towards DevOps of Zero-Knowledge Proofs on Blockchain: LLM-Enhanced Proof Generation and Contract Deployment},
  author={Ismayilov, Goshgar Can},
  booktitle={2025 International Conference on Big Data, Knowledge and Control Systems Engineering (BdKCSE)},
  pages={1--8},
  year={2025},
  organization={IEEE}
}

@ARTICLE{zkstarks,
  title={Scalable, transparent, and post-quantum secure computational integrity},
  author={Ben-Sasson, Eli and Bentov, Iddo and Horesh, Yinon and Riabzev, Michael},
  journal={Cryptology ePrint Archive},
  year={2018}
}

@CONFERENCE{zksnarks,
  title={Succinct $\{$Non-Interactive$\}$ zero knowledge for a von neumann architecture},
  author={Ben-Sasson, Eli and Chiesa, Alessandro and Tromer, Eran and Virza, Madars},
  booktitle={23rd USENIX Security Symposium (USENIX Security 14)},
  pages={781--796},
  year={2014}
}

@CONFERENCE{bulletproofs,
  title={Bulletproofs: Short proofs for confidential transactions and more},
  author={B{\"u}nz, Benedikt and Bootle, Jonathan and Boneh, Dan and Poelstra, Andrew and Wuille, Pieter and Maxwell, Greg},
  booktitle={2018 IEEE symposium on security and privacy (SP)},
  pages={315--334},
  year={2018},
  organization={IEEE}
}

@ARTICLE{plonk,
  title={Plonk: Permutations over lagrange-bases for oecumenical noninteractive arguments of knowledge},
  author={Gabizon, Ariel and Williamson, Zachary J and Ciobotaru, Oana},
  journal={Cryptology ePrint Archive},
  year={2019}
}

@CONFERENCE{cao2022,
  title={Hybrid smart contracts for privacy-preserving-aware insurance compensation},
  author={Cao, Sheng and Zhang, Qian and Wang, Dongdong and Xiangli, Peng and Zhang, Xiaosong},
  booktitle={2022 IEEE Wireless Communications and Networking Conference (WCNC)},
  pages={1533--1538},
  year={2022},
  organization={IEEE}
}

@ARTICLE{ismayilov20252,
  title={PTTS: Zero-knowledge proof-based private token transfer system on Ethereum blockchain and its network flow based balance range privacy attack analysis},
  author={Ismayilov, Goshgar and {\"O}zturan, Can},
  journal={Journal of Network and Computer Applications},
  volume={233},
  pages={104045},
  year={2025},
  publisher={Elsevier}
}

@SOFTWARE{gpt4o,
  title={$\mbox{GPT-4o}$ Model},
  author={OpenAI},
  url={https://openai.com/index/hello-gpt-4o/},
  note={Accessed: 2026-06-25},
  year={2024}
}

@SOFTWARE{metamask,
  title={MetaMask Wallet},
  author={MetaMask},
  url={https://metamask.io/},
  note={Accessed: 2026-06-25},
  year={2016}
}

@ARTICLE{kondyrev2024,
  title={Comparative Efficiency Analysis of Hashing Algorithms for Use in zk-SNARK Circuits in Distributed Ledgers},
  author={Kondyrev, DO},
  journal={Programming and Computer Software},
  volume={50},
  number={4},
  pages={283--291},
  year={2024},
  publisher={Springer}
}

@ARTICLE{gu2024,
  title={zk-oracle: Trusted off-chain compute and storage for decentralized applications},
  author={Gu, Binbin and Nawab, Faisal},
  journal={Distributed and Parallel Databases},
  volume={42},
  number={4},
  pages={525--548},
  year={2024},
  publisher={Springer}
}

@CONFERENCE{grassi2021,
  title={Poseidon: A new hash function for $\{$Zero-Knowledge$\}$ proof systems},
  author={Grassi, Lorenzo and Khovratovich, Dmitry and Rechberger, Christian and Roy, Arnab and Schofnegger, Markus},
  booktitle={30th USENIX Security Symposium (USENIX Security 21)},
  pages={519--535},
  year={2021}
}

@CONFERENCE{wu2023,
  title={Smart contract-based E-voting system using homomorphic encryption and zero-knowledge proof},
  author={Wu, Yuxiao and Kasahara, Shoji},
  booktitle={International Conference on Applied Cryptography and Network Security},
  pages={67--83},
  year={2023},
  organization={Springer}
}

@ARTICLE{panja2018,
  title={A secure end-to-end verifiable e-voting system using zero knowledge based blockchain},
  author={Panja, Somnath and Roy, Bimal Kumar},
  journal={Cryptology ePrint Archive},
  year={2018}
}

@ARTICLE{miao2023,
  title={Secure and privacy-preserving voting system using zero-knowledge proofs},
  author={Miao, Yizhuo},
  journal={Applied and Computational Engineering},
  volume={8},
  number={1},
  pages={328--333},
  year={2023}
}

@ARTICLE{heiss2023,
  title={Verifiable carbon accounting in supply chains},
  author={Heiss, Jonathan and Oegel, Tahir and Shakeri, Mehran and Tai, Stefan},
  journal={IEEE Transactions on Services Computing},
  volume={17},
  number={4},
  pages={1861--1874},
  year={2023},
  publisher={IEEE}
}

@ARTICLE{itanyi2023,
  title={A car insurance claim processing prototype using smart contracts and zero-knowledge proofs},
  author={Itanyi, Mamudu Francis and Modi, B and Friday, Mamudu},
  journal={Dutse Journal of Pure and Applied Sciences},
  volume={9},
  number={4b},
  pages={362--371},
  year={2023}
}

@SOFTWARE{yahoo_breach,
  author = {Wikipedia},
  title = {Yahoo data breaches},
  url = {https://en.wikipedia.org/wiki/Yahoo_data_breaches},
  note={Accessed: 2026-06-25},
  year = {2014},
}

@ARTICLE{guo2025,
  title={Trusted Execution Environments for Blockchain: Towards Robust, Private, and Scalable Distributed Ledgers},
  author={Guo, Zhikang and Pan, Heng and He, Ang and Dai, Yueyue and Huang, Xiaoyan and Si, Xueming and Yuen, Chau and Zhang, Yan},
  journal={IEEE Internet of Things Journal},
  year={2025},
  publisher={IEEE}
}

@ARTICLE{sun2021,
  title={A survey on zero-knowledge proof in blockchain},
  author={Sun, Xiaoqiang and Yu, F Richard and Zhang, Peng and Sun, Zhiwei and Xie, Weixin and Peng, Xiang},
  journal={IEEE network},
  volume={35},
  number={4},
  pages={198--205},
  year={2021},
  publisher={IEEE}
}

@ARTICLE{zhou2021,
  title={Using secure multi-party computation to protect privacy on a permissioned blockchain},
  author={Zhou, Jiapeng and Feng, Yuxiang and Wang, Zhenyu and Guo, Danyi},
  journal={Sensors},
  volume={21},
  number={4},
  pages={1540},
  year={2021},
  publisher={MDPI}
}

@ARTICLE{zhou2024,
  title={Efficient zero-knowledge range arguments and privacy-preserving applications},
  author={Zhou, Yue},
  journal={The Australian National University},
  year={2024}
}

@CONFERENCE{gokulakrishnan2025,
  title={Scalable Supply Chain Product Source Verification Using Zero-Knowledge Proofs},
  author={Gokulakrishnan, D and Sinha, Tanya and others},
  booktitle={2025 International Conference on Computing and Communication Technologies (ICCCT)},
  pages={1--5},
  year={2025},
  organization={IEEE}
}

@ARTICLE{de2022,
  title={Leveraging self-sovereign identity, blockchain, and zero-knowledge proof to build a privacy-preserving vaccination pass},
  author={de Vasconcelos Barros, Mauricio and Schardong, Frederico and Felipe Cust{\'o}dio, Ricardo},
  journal={Blockchain, and Zero-Knowledge Proof to Build a Privacy-Preserving Vaccination Pass},
  year={2022}
}

@ARTICLE{gantait2022,
  title={Zero knowledge identification and verification of voting systems},
  author={Gantait, Arunava and Goyal, Rajit and Rizvi, Syed Sajid Husain and Haram, Zaira},
  journal={arXiv preprint arXiv:2212.06388},
  year={2022}
}

@ARTICLE{kuszmaul2019,
  title={Verkle trees},
  author={Kuszmaul, John},
  journal={Verkle trees},
  volume={1},
  number={1},
  pages={1--10},
  year={2019}
}

\appendix

\end{document}